\documentclass[aps,prl,twocolumn,groupedaddress,amsfonts,showpacs]{revtex4}
\usepackage{graphicx,color}
\usepackage{bm}
\usepackage[latin1]{inputenc}

\begin{document}

\title{All electrical measurement of the density of states in (Ga,Mn)As}

\author{D. Neumaier}
\email{daniel.neumaier@physik.uni-regensburg.de}
%\homepage{http://www.physik.uni-regensburg.de/forschung/weiss/}
\author{M. Turek}
\altaffiliation[Present address: ]{Fraunhofer ISE , Halle, Germany}
\author{U. Wurstbauer}
\altaffiliation[Present address: ]{Institut f\"{u}r Angewandte
Physik, University of Hamburg, Germany}
\author{A. Vogl}
\author{M. Utz}
\author{W. Wegscheider}
\author{D. Weiss}

\affiliation{Institut für Experimentelle und Angewandte Physik,
University of Regensburg, 93040 Regensburg, Germany}

\date{\today}

\begin{abstract}
We report on electrical measurements of the effective density of
states in the ferromagnetic semiconductor material (Ga,Mn)As. By
analyzing the conductivity correction due to enhanced
electron-electron interaction the electrical diffusion constant was
extracted for (Ga,Mn)As samples of different dimensionality. Using
the Einstein relation allows to deduce the effective density of
states of (Ga,Mn)As at the Fermi energy.

\end{abstract}

\pacs{71.20.-b, 75.50.Pp, 73.23.-b}%
\keywords{}

\maketitle

The ferromagnetic semiconductor (Ga,Mn)As \cite{Ohno} has been
studied intensely over the last decade and has become a model system
for future spintronics applications \cite{Fabian,Fabian2}. With
typical Mn-concentrations between 1 \% and 15 \% maximum Curie
temperatures of up to $\sim180$ K have been reported
\cite{Olejnik,Wang}. Mn atoms on Ga-sites provide both holes and
magnetic moments. For Mn concentrations larger than 1 \% the
impurity wavefunctions at the Fermi energy overlap and a metallic
state forms. The ferromagnetic order between the magnetic moments of
the Mn-ions is mediated by the delocalized holes \cite{Dietl}. A
topic of current debate is whether the holes reside in an impurity
band, detached and above the valence band or in the valence band
\cite{Jungwirth}. A mean field picture based on the latter scenario
allowed to predict, e.g. Curie temperature \cite{Dietl} or
magnetocrystalline anisotropies \cite{Sawicki} in (Ga,Mn)As
correctly. On the other hand optical absorption experiments, carried
out, e.g. in Ref. \cite{Burch,Ando}, suggest that even for high
manganese concentrations of up to 7 \% the Fermi energy stays in an
impurity band, detached from the valence band, with a high effective
hole mass of order ten free electron masses $m_e$ \cite{Burch}.
However, there is also indication that the impurity band and the
valence band have completely merged as discussed in Ref.
\cite{Jungwirth} and references therein. In the present letter we
make use of the well known quantum mechanical conductivity
correction due to electron-electron interaction (EEI) to extract the
diffusion constant and hence the density of states at the Fermi
energy, $N(E_F)$. The electrically measured values of $N(E_F)$ will
be compared with recent theoretical calculations.

In ferromagnetic (Ga,Mn)As the conductivity is decreasing with
decreasing temperature below 10 K. This conductivity decrease can be
explained by enhanced electron-electron interaction \cite{EEI}. The
effect of EEI arises from a modified screening of the
Coulomb-potential due to the carriers' diffusive  motion and depends
on the dimensionality of the conductor \cite{Lee}. As the
conductivity decrease due to enhanced electron-electron interaction
is depending on the electrical diffusion constant $D$, a detailed
analysis of the conductivity decrease, different for different
dimensionality, provides experimental access to the diffusion
constant. Using the Einstein relation $\sigma=N(E_F)D\mathrm{e}^2$,
with the conductivity $\sigma$, the effective density of states at
Fermi's energy, $N(E_F)$, can be determined.

\begin{table}
\begin{tabular}{|l|l|l|l|l|l|l|}
\hline
Sample &$l$ ($\mu$m)&$w$ ($\mu$m)&$t$ (nm)&$N$&$T_C$ (K)&p ($10^{26}$ /m$^3$)\\
\hline
1$_{1D}$&7.5&0.042&42&25&90&3.8\\
1$_{1D}$A&7.5&0.042&42&25&150&9.3\\
2$_{1D}$&7.5&0.035&42&12&90&3.8\\
1$_{1D-2D}$A&10&0.067&30&25&150&8.6\\
2$_{1D-2D}$A&10&0.092&30&25&150&8.6\\
3$_{1D-2D}$A&10&0.170&30&25&150&8.6\\
4$_{1D-2D}$A&10&0.242&30&25&150&8.6\\
1$_{2D}$&180&11&42&1&90&3.8\\
1$_{3D}$&240&10&150&1&?&1.4\\
2$_{3D}$&240&10&300&1&75&2.1\\
\hline
\end{tabular}
\caption{Length \emph{l}, width \emph{w}, thickness \emph{t} and
number of lines parallel \emph{N} of the samples. Curie temperature
$T_C$ and carrier concentration \emph{p} were taken on reference
samples from the corresponding wafers. Annealed samples are marked
by "A".} \label{sampleparameter}
\end{table}

To investigate electron-electron interaction in quasi 1D, 2D and 3D
systems we fabricated Hall-bar mesas (2D and 3D) and wire arrays (1D
and crossover regime from 1D to 2D) out of several wafers, having a
(Ga$_{1-x}$,Mn$_x$)As layer on top of semi-insulating GaAs. The
nominal Mn-concentration $x$ was approx. 4 \% (sample 1$_{3D}$ and
2$_{3D}$) and $\sim6$ \% (other samples). The relevant parameters of
the samples are listed in table \ref{sampleparameter}. The
dimensionality for EEI is defined by the number of spatial
dimensions larger than the thermal diffusion length $L_T=\sqrt{\hbar
D/k_BT}$. In (Ga,Mn)As $L_T\approx120-200$ nm at 20 mK, depending on
the exact value of the diffusion constant $D$. Hence the thick
Hall-bar mesas (150 nm and 300 nm) can be considered as quasi 3D,
while the thin Hall-bar mesa (42 nm) is quasi 2D, at least below
$\sim500$ mK. The smallest wires (42 nm and 35 nm) behave quasi 1D
and the wider wires (62nm to 242 nm) are  in the crossover regime
from 1D to 2D, as is shown below. Arrays of wires with $N$ wires in
parallel were fabricated to suppress universal conductance
fluctuations by ensemble averaging. The Hall-bars were fabricated
using optical lithography and wet chemical etching. For fabricating
the wire arrays we used electron-beam lithography and chemical dry
etching. The contact pads to the devices were made by thermal
evaporation of Au and lift-off. The measurements of the conductivity
were performed in a top-loading dilution refrigerator using standard
four-probe lock-in technique. To avoid heating of the charge
carriers small measuring currents (25 pA to 4 nA, depending on the
sample's resistance) and proper shielding were crucial. For each
sample the measuring current was kept fixed for all temperatures. To
suppress conductivity contributions due to weak localization we
applied a perpendicular magnetic field of $B=3$ T. At $B=3$ T no
weak localization can be observed in (Ga,Mn)As \cite{WAL,Rozkinson}
even at 20 mK.

\begin{figure}
\includegraphics[width=\linewidth]{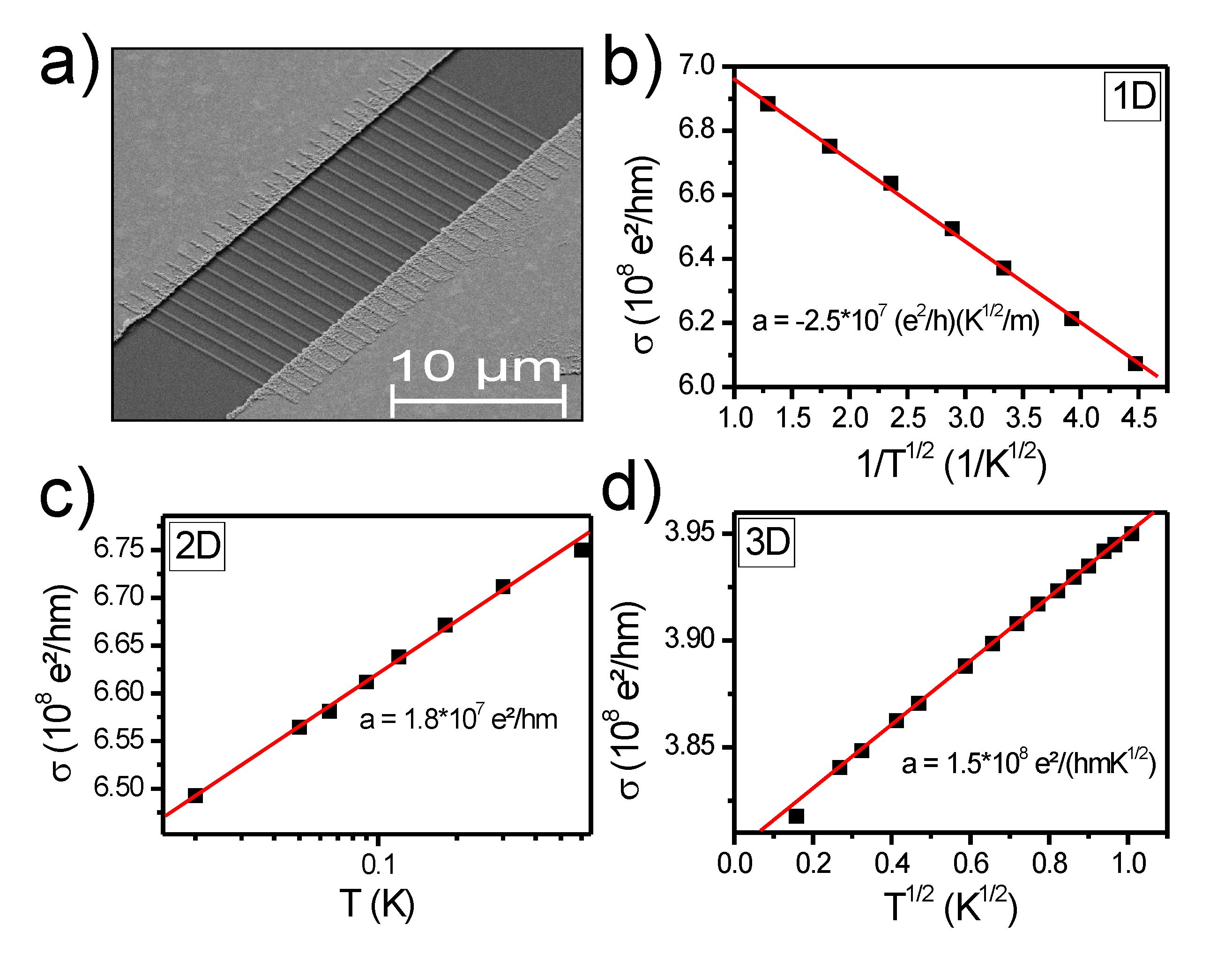}
\caption{a) Electron micrograph of a line array having 25 lines in
parallel (Sample 2$_{1D-2D}$A). The width of the lines is 92 nm, the
length is 10 $\mu$m. b), c) and d): Conductivity of the quasi 1D
line-array 1$_{1D}$ (b), the quasi 2D Hall-bar 1$_{2D}$ (c) and the
quasi 3D Hall-bar 2$_{3D}$ (d) plotted versus temperature. The
straight lines are guide for the eyes. The slope of the lines are
given.}
\end{figure}

According to Lee and Ramakrishnan \cite{Lee} the temperature
dependency of the conductivity correction due to EEI is depending on
the dimensionality of the sample with respect to $L_T$. For
1D-systems the expected temperature dependency is $\propto
-1/\sqrt{T}$, for 2D $\propto \mathrm{log}_{10}(T/T_0)$ and for 3D
$\propto\sqrt{T}$.  Corresponding data for 1D, 2D and 3D
(Ga,Mn)As-samples, shown in figure 1b)-d),  confirm the expected
temperature dependency below 1 K. Hence the decreasing conductance
with decreasing temperature in (Ga,Mn)As can be attributed to EEI.

The size of the conductivity correction due to electron-electron
interaction is depending on the diffusion constant $D$ in 1D-systems
\cite{Lee}:
\begin{equation}
\Delta \sigma=-\frac{F^{1D}}{\pi
wt}\frac{e^2}{\hbar}\sqrt{\frac{\hbar D}{k_BT}},
\end{equation}
and also in 3D-systems \cite{Lee}:
\begin{equation}
\Delta
\sigma=\frac{F^{3D}}{4\pi^2}\frac{e^2}{\hbar}\sqrt{\frac{k_BT}{\hbar
D}}.
\end{equation}

As the conductivity correction due to EEI is also depending on the
screening parameters $F^{1D,2D,3D}$ one needs to know the
corresponding parameter to extract $D$ from the conductivity
correction. Only in quasi 2D-systems the conductivity correction is
independent on the diffusion constant \cite{Lee}:
\begin{equation}
\Delta
\sigma=\frac{F^{2D}}{2t\pi^2}\frac{e^2}{\hbar}\log{\frac{T}{T_0}}.
\end{equation}
Hence in the 2D case $F^{2D}$ can be directly extracted from
experiment. As already shown in previous work \cite{EEI} the
screening parameter $F^{2D}$ in (Ga,Mn)As ranges from 1.8 to 2.6 and
is in excellent agreement with the screening parameter in Co,
Co/Pt-multilayers and Permalloy. In these ferromagnetic metals
$F^{2D}$ is between 2.0 and 2.6
\cite{BrandsCo,BrandsCo2,BrandsCoPt,Py}. Thus using the well known
parameters  $F^{1D}$ of other ferromagnetic metals is a good
approximation for $F^{1D}$ in (Ga,Mn)As. In Ni and Py nanowires
$F^{1D}$ is 0.83 and 0.77 respectively \cite{Ono,Py}. Consequently
$F^{1D}=0.80\pm12$ should be a good approximation for the screening
parameter of quasi 1D (Ga,Mn)As samples. With this $F^{1D}$
parameter we can calculate the diffusion constant of sample 1$_{1D}$
using equation (1): $D=10\pm3\cdot10^{-5}$ m$^2$/s. Using the
Einstein relation this value corresponds to an effective density of
states $N(E_F)=1.1\pm0.3\cdot10^{46}$ /Jm$^3$ at the Fermi energy.
In figure 3 $N(E_F)$ is plotted versus the carrier concentration
(green squares) for all investigated 1D-samples. Here, the
uncertainty in determining the carrier concentration is $\sim$10 \%.

It is more difficult to estimate the value of $F^{3D}$ as no data
are available for 3D ferromagnetic metals. Therefore we have to rely
on theoretical predictions for the screening parameter: $F^{3D}=1.2$
\cite{Schwab}. Though the calculations of the 2D screening parameter
($F^{2D}=2.3$ \cite{Schwab}) agree well with the experimental values
of different ferromagnets ($F^{2D}=1.8...2.6$), they are less
accurate for 1D systems. For 1D systems $F^{1D}$ was calculated to
be 1.6, while the typical experimental values of $F^{1D}\approx0.8$
are by a factor of 2 smaller \cite{Py,Ono}.

Hence by using the theoretical value for $F^{3D}$ we need to take
into account an uncertainty of order ~100 \%. Using equation (2) and
the theoretical value for $F^{3D}=1.2$ \cite{Schwab} we arrive at
$D=2.2\cdot10^{-5}$ m$^2$/s for sample 2$_{3D}$. Using the Einstein
relation this corresponds to $N(E_F)=2.0\cdot10^{46}$ /Jm$^3$, with
a high uncertainty of approx. 200 \%, as $D$ is depending
quadratically on $F^{3D}$.

\begin{figure}
\includegraphics[width=\linewidth]{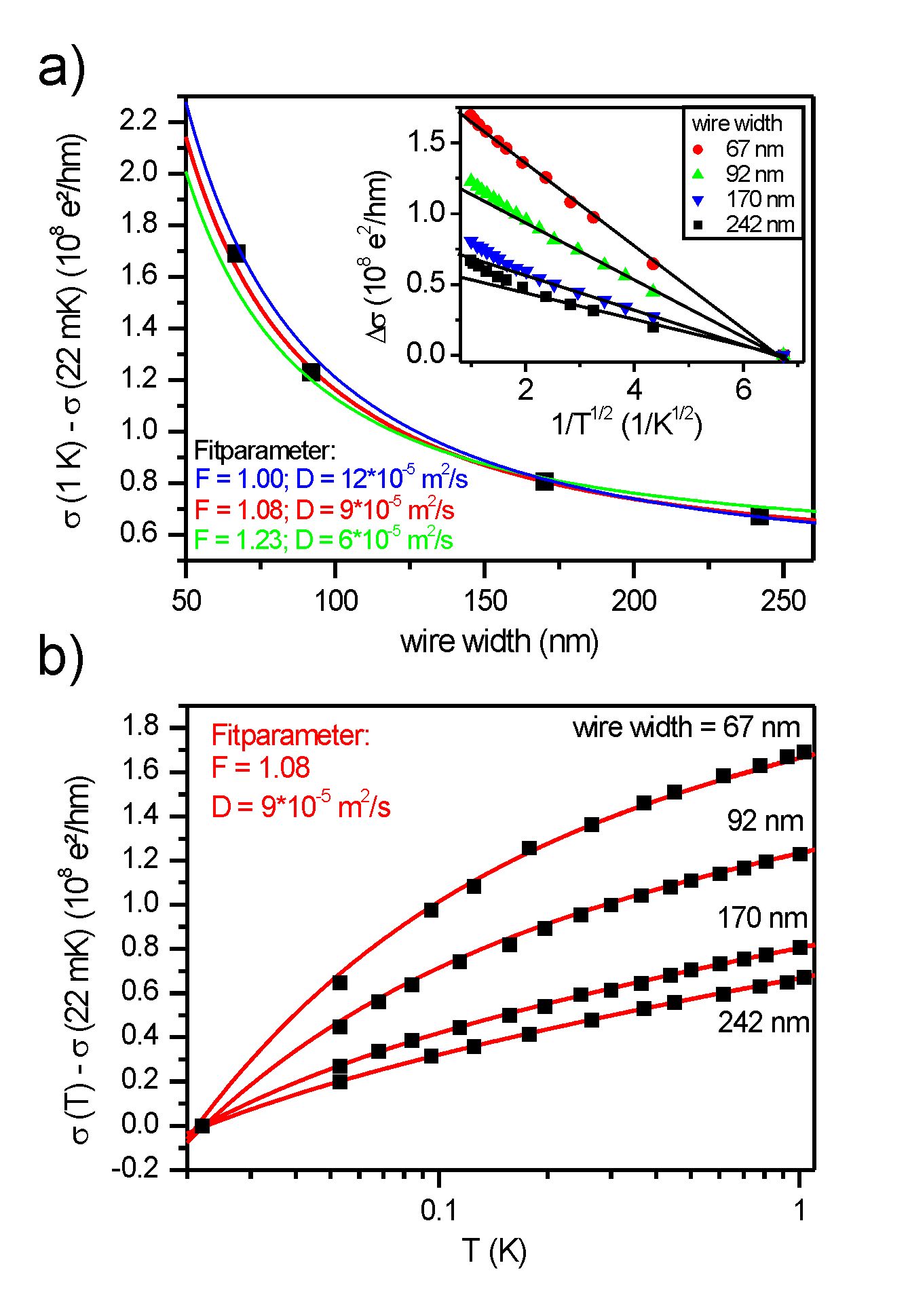}
\caption{a) Conductivity change from 1 K to 22 mK of four line
arrays (sample 1$_{1D-2D}$A...4$_{1D-2D}$A) plotted versus the wire
width. The red line is the best fit of the data to equation
(\ref{1D2D}). The blue and green lines are fits using equation
(\ref{1D2D}) and a diffusion constant of $12\cdot10^{-5}$ m$^2$/s
and $6\cdot10^{-5}$ m$^2$/s respectively. In the inset the
conductivity change of four wire arrays (sample
1$_{1D-2D}$A...4$_{1D-2D}$A) is plotted versus $1/\sqrt{T}$. The
slopes are guide for the eyes. b) Conductivity change of the four
line arrays (sample 1$_{1D-2D}$A...4$_{1D-2D}$A) plotted versus
temperature. The red lines are calculated using equation (4) and the
parameters obtained by fitting the data in a).}
\end{figure}

The different temperature dependence of EEI in 1D
($\Delta\sigma\propto -1/\sqrt{T}$) and 2D ($\Delta\sigma\propto
\mathrm{ln}{T}$) together with the temperature dependence of the
thermal length $L_T=\sqrt{\hbar D/k_BT}$ allows to use another
scheme to extract the diffusion constant and hence $N(E_F)$. By
measuring the dimensional crossover, i.e. the change of the
temperature dependence of the conductivity correction as a function
of the sample size one can fit both, diffusion constant and
screening parameter independently. Here we used the crossover from
1D to 2D to determine $D$.  For this experiment wire arrays with
wire widths, ranging from 62 nm to 242 nm (sample
1$_{1D-2D}$A...4$_{1D-2D}$A in table 1), were patterned on the same
wafer. These four wires are in the crossover regime between 1D and
2D. This is most clearly seen in the inset of figure 2a and in
figure 2b. In Fig. 2b the logarithmical temperature dependency,
expected for 2D EEI, is only describing the widest wire at high
temperatures satisfactorily, while the $1/\sqrt{T}$ dependency,
expected for 1D EEI, only holds for the smallest wire at low
temperatures in the inset of Fig. 2a. In the crossover regime the
conductivity correction due to EEI is given by an interpolation
formula \cite{Neuttiens}:
\begin{eqnarray}
\Delta\sigma t=-F\frac{\mathrm{e}^2}{\pi\hbar}\sum_{n=0}^\infty\left[\frac{w^2}{L^2_T}+(n\pi)^2\right]^{-1/2}-\nonumber \\
\left[\frac{w^2}{L^2_{T_0}}+(n\pi)^2\right]^{-1/2},
\label{1D2D}
\end{eqnarray}
with one screening parameter $F$ and $T_0$, the lowest temperature.
Figure 2a shows the conductivity change from 1 K to 22 mK of all
four wire arrays. The conductivity change increases markedly with
decreasing wire width. To extract the characteristic parameters we
fitted the data using equation (4) with $D$ and $F$ as free
parameters. The diffusion constant affects essentially the width
dependence (x-scale)  while the screening parameter $F$ shifts the
curve on the y-scale. Hence the fit is unique and allows to extract
$D$ and $F$ independently. The best fitting result was obtained by
using $D=9\cdot10^{-5}$ m$^2$/s and $F=1.08$ (red line). To
illustrate the sensitivity of the fitting procedure on $D$ we also
plotted equation (4) using $D=12\cdot10^{-5}$ m$^2$/s (blue line)
and $D=6\cdot10^{-5}$ m$^2$/s (green line). Here $F$ was the free
parameter to optimize the fit. Both traces (blue and green) describe
the experimental data less satisfying than the red trace. Hence the
measurement of the dimensional crossover from 1D to 2D results in
$D=9\pm1.5\cdot10^{-5}$ m$^2$/s. In figure 2b the conductivity
change with respect to 22 mK is plotted for all four wire arrays
(sample 1$_{1D-2D}$A to 4$_{1D-2D}$A) versus temperature. The red
lines are the calculated conductivity correction given by equation
(4). The parameters used were  $D=9\cdot10^{-5}$ m$^2$/s and
$F=1.08$. The conductivity correction in the whole temperature range
from 22 mK to 1 K of all four wire arrays is perfectly described by
using only these two parameters $D$ and $F$. Also in the crossover
regime from 1D to 2D the observed screening parameter $F=1.08$ is in
excellent agreement with the screening parameter observed in Co
($F=0.95$) \cite{BrandsCo}. This underlines the universal character
of the conductivity correction due to EEI. From the obtained
diffusion constant we can estimate the effective density of states
using the Einstein relation: $N(E_F)=1.6\pm0.3\cdot10^{46}$ /Jm$^3$.

To check the consistency of both presented methods, we can also
treat the four samples (1$_{1D-2D}$A to 4$_{1D-2D}$A) as quasi 1D at
low temperatures (indicated by the straight lines in the inset of
figure 3a) and calculate the diffusion constant using equation (1)
and $F^{1D}=0.8$ as described above. When doing so we obtain
$D=8.4\pm2.5\cdot10^{-5}$ m$^2$/s. This is in good agreement with
the value estimated by fitting the crossover from 1D to 2D and hence
the two methods are consistent.

\begin{figure}
\includegraphics[width=\linewidth]{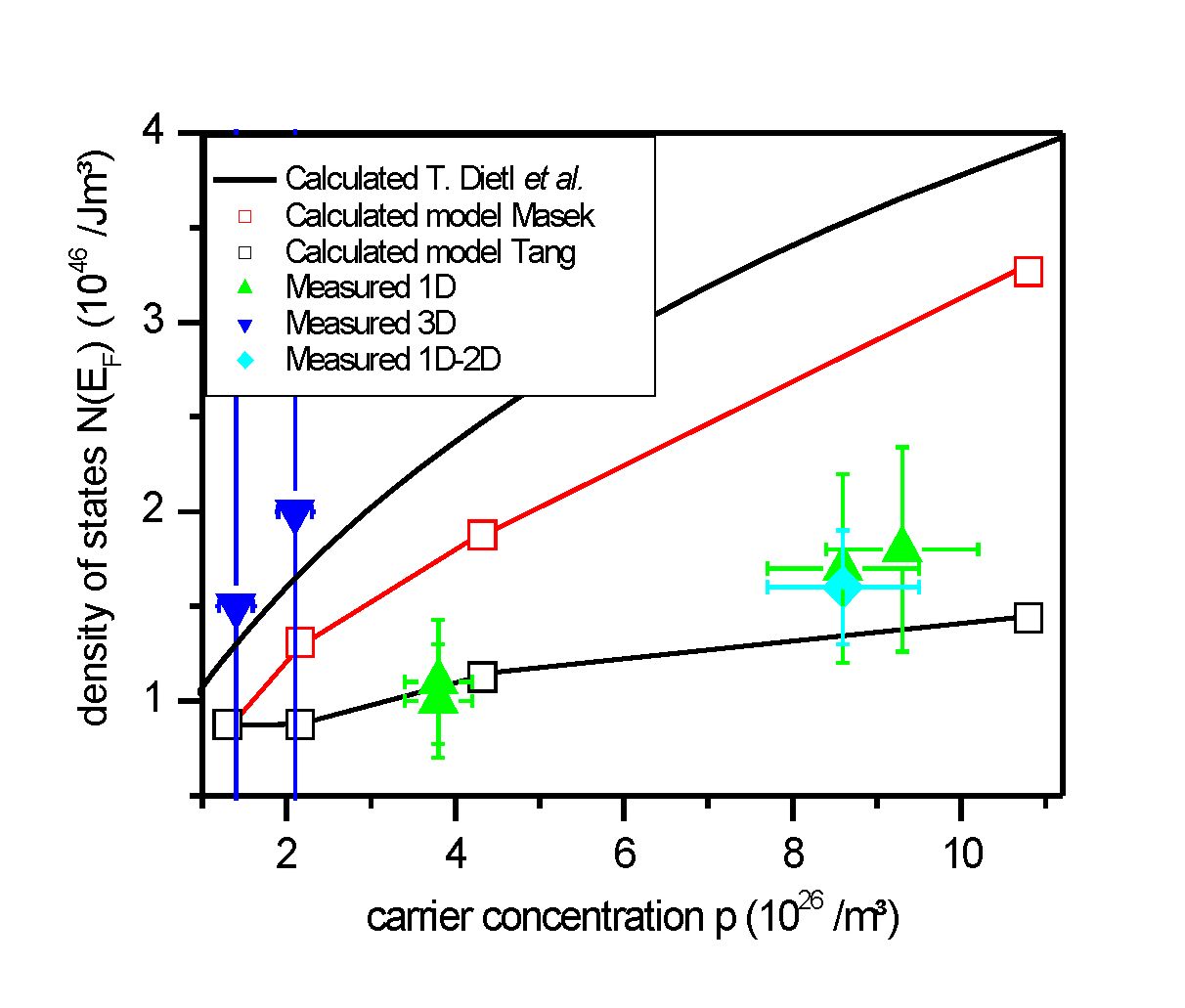}
\caption{Density of states at Fermi's energy plotted versus carrier
concentrations. The squares are calculated using the model of Masek
\textit{et al.}  and the model of Tang and Flattè respectively. The
solid line gives $N(E_F)$ calculated using a $6\times6$ $k\cdot p$
model with parameters for the GaAs valence band \cite{Dietl2}. The
filled symbols are measured using EEI in 1D (green triangle), 3D
(blue triangle) and in the crossover regime from 1D to 2D (blue
diamond).}
\end{figure}

Figure 3 summarizes our findings and shows the extracted effective
density of states versus the carrier concentration (filled symbols).
The data are compared to calculations based on three different
models: The solid line stems from a $6\times6$ $k\cdot p$ model with
parameters for the GaAs valence band, calculated by T. Dietl
\textit{et al.} and based on reference \cite{Dietl2}. The
corresponding effective hole mass is $\sim1m_e$ in the investigated
range of carrier concentration. In addition we performed numerical
simulations which are based on a multi-band tight-binding approach
applied to disordered bulk systems using two different parameter
sets. The first model was derived from first principles calculations
for (Ga,Mn)As \cite{Masek} (model Masek). The second one describes
the Mn impurities by a modified on-site potential and a
spin-dependent potential at the four nearest neighbor As sites which
reproduce the experimental binding energy of 113 meV \cite{Tang}
(model Tang). A detailed description of the method and the two
models is given in reference \cite{Turek}. Neither model exhibits a
detached impurity band for Mn concentrations larger than 1\%. The
effective masses were estimated to lie in the range $m^*=0.4...0.6
m_e$ for the considered carrier concentrations with only minor
quantitative differences between the two models. Due to the
increasing number of holes the Fermi energy moves deeper into the
valence band with increasing disorder. The experimental data for the
density of states lie between the predicted values of the models.
Although the data seem to be closer to the model of Tang and Flattè,
the relative high uncertainty of the experimental results does not
allow to give a definite answer which model describes the
experimental results better. For the case that Fermi's energy lies
within a detached impurity band no calculations for the density of
states are available. However assuming a parabolic band, one finds
that $N(E_F)\propto m^*$, for a given carrier concentration. Hence
an effective mass $m^*>>m_e$, as expected for a detached impurity
band, leads to an effective density of states at Fermi's energy far
away from the measured values.

In conclusion we have demonstrated that the effective density of
states at the Fermi energy of (Ga,Mn)As can be extracted from
conductivity measurements, i.e. an analysis of the conductivity
correction due to EEI. The measured values of $N(E_F)$ are
consistent with a picture that the Fermi energy is located within
the GaAs valence band or an impurity band, merged with the valence
band. Our experimental finding with effective masses of $\sim1m_e$
is however at odds with a detached impurity band, where the
effective hole mass is much larger than $m_e$.

Acknowledgement: We thank T. Dietl and J. Fabian for stimulating
discussions and the Deutsche Forschungsgemeinschaft (DFG) for their
financial support via SFB 689.

%\begin{references}

% \end{multicols}

\end{document}